\documentclass[]{qjmam}
\usepackage{amssymb, amsmath, bm, color}

\startpage{1}
\yr{2018}
\vol{52}
\issue{1}

\DeclareMathAlphabet\mathbit
    \encodingdefault\rmdefault\bfdefault\itdefault
\DeclareOldFontCommand{\bi}{\normalfont\bfseries\itshape}{\mathbit}

\newcommand{\be}{\begin{equation}}
\newcommand{\ee}{\end{equation}}

\def\fakebold#1{\relax\ifvmode\leavevmode\fi%
\ifmmode%
\setbox0=\hbox{$#1$}%
\else%
\setbox0=\hbox{#1}%
\fi%
\kern-.02em\copy0 \kern-\wd0%
\kern .04em\copy0 \kern-\wd0%
\kern-.0125em\raise.02em\box0%
}%

\renewcommand{\leq}{\leqslant}

\usepackage[numbers]{natbib}
\newcommand{\figpath}{}

\newcommand{\pfrac}[2]{
\frac{\partial #1}{\partial #2}
}

\begin{document}

\title[asymptotics~of~self-similar~blow-up~profiles~of~the~thin~film~equation]{MATCHED ASYMPTOTIC ANALYSIS OF SELF-SIMILAR BLOW-UP PROFILES OF THE THIN FILM EQUATION}

\author[M.~C. Dallaston] {MICHAEL. C. DALLASTON}

\address{School of Computing, Engineering and Mathematics, Coventry University,\\
Coventry, {\rm CV1 5FB}, United Kingdom}

\received{\recd \today. \revd XXX}


\maketitle

\eqnobysec

\begin{abstract} 
We consider asymptotically self-similar blow-up profiles of the thin film equation consisting of a stabilising fourth order and destabilising second order term. It has previously been shown that blow up is only possible when the exponent in the second order term is above a certain critical value (dependent on the exponent in the fourth order term). We show that in the limit that the critical value is approached from above, the primary branch of similarity profiles exhibits a well-defined structure consisting of a peak near the origin, and a thin, algebraically decaying tail, connected by an inner region equivalent (to leading order) to a generalised version of the Landau--Levich `drag-out' problem in lubrication flow. Matching between the regions ultimately gives the asymptotic relationship between a parameter representing the height of the peak and the distance from the criticality threshold.  The asymptotic results are supported by numerical computations found using continuation.
\end{abstract}


\section{Introduction}

In this paper we consider the partial differential equation
\begin{equation}
\label{eq:pde}
\pfrac{h}{t} + \pfrac{}{x}\left[ h^m \pfrac{^3h}{x^3} + h^n \pfrac{h}{x} \right] = 0,
\end{equation}
which appears extensively in lubrication (or thin film) theory in a variety of contexts \cite{Oron1997,Craster2009}.  The equation (\ref{eq:pde}) may arise as a model of the thickness $h$ of a liquid film over space $x$ and time $t$, where the film is stabilised by a fourth order term (usually coming from a surface tension term on a free boundary) and destabilised by a second order term, which may come from including one of a wide range of physical effects that may be of interest, for example intermolecular forces, gravity, thermocapillarity, or inertia.  (\ref{eq:pde}) has been nondimensionalised such that the only parameters that appear are the two exponents $m$ and $n$ in the stabilising and destabilising terms, respectively.  The implicit scales in the initial condition and domain size don't play a role in self-similar singularity formation, which is a purely local phenomenon.

When (\ref{eq:pde}) models a thin film sheet bounded by a free interface on one side and a solid substrate on the other, the value for $m=3$ arises from the lubrication approximation.  In this context many values of $n$ have been considered, depending on the nature of the force destabilising the interface.  Some phenomona that fit into this framework are van der Waals forces ($n=-1$ \cite{Zhang1999,Witelski1999}), Rayleigh--Taylor (gravity-driven) instability ($n=3$ \cite{Yiantsios1989}), thermocapillary forces and vapour recoil from evaporation ($n=0$--$2$ and $n=0$--$3$, respectively, depending on the importance of nonequilibrium thermal effects \cite{Boos1999, Burelbach1988}), and inertia of an inclined or vertical film ($n=6$, \cite{Pumir1983}).  Many such phenomena are detailed in the review papers \cite{Oron1997, Craster2009}.  If one considers a thin filament in a Hele--Shaw cell, then $m=1$ \cite{Almgren1996a,Almgren1996b}.  The value $m=0$ appears in the analysis of limiting behaviour of the Cahn--Hilliard equation \cite{Novick1984,Elliott1986,Evans2006}, and has also recently arisen in the near-rupture behaviour of two fluid layer systems bounded between two walls \cite{Zhao2018}.

Behaviour of equations such as (\ref{eq:pde}) near a singularity is often self-similar in nature, and a great deal of progress is made by assuming self-similarity.  
In general, there are often discrete families of self-similar solutions to nonlinear partial differential equations, and a full understanding of the time-dependent dynamics requires the analysis of the linear stability of such solutions in a coordinate system in which a self-similar profile corresponds to a steady state \cite{Witelski2000,Bernoff2002,Bernoff2010}.
A number of computational and analytic studies have examined the self-similar finite time singular behaviour of (\ref{eq:pde}) in particular for different values of ranges of $m$ and $n$.  Generally speaking, when $n$ is sufficiently small, the generic behaviour of (\ref{eq:pde}) is to rupture at finite time, that is, $h(x_0,t) \to 0$ as $t\to t_0^-$ for some singular point and time $x_0$ and $t_0$ \cite{Zhang1999,Witelski1999,Dallaston2017,Dallaston2018}.  For given $m$, the exact value of $n$ above which rupture  cannot occur is an open problem; for $m=3$, \cite{Dallaston2018} found that the stability of the primary branch of self-similar rupture solutions is first lost via a pair of Hopf bifurcations, leading to the presence of `discrete' self similar solutions, in which profiles repeat on discrete, geometrically shrinking scales.  
Self-similar analysis has also been applied to related problems, such as source-type solutions (with no finite time singularity) \cite{Evans2007a}, the thin film equation corresponding to (\ref{eq:pde}) without a second order term \cite{Bernis2000}, and sixth-order analogues of (\ref{eq:pde}) \cite{Evans2007b,Evans2007c}.

\begin{figure}
\centering
\includegraphics{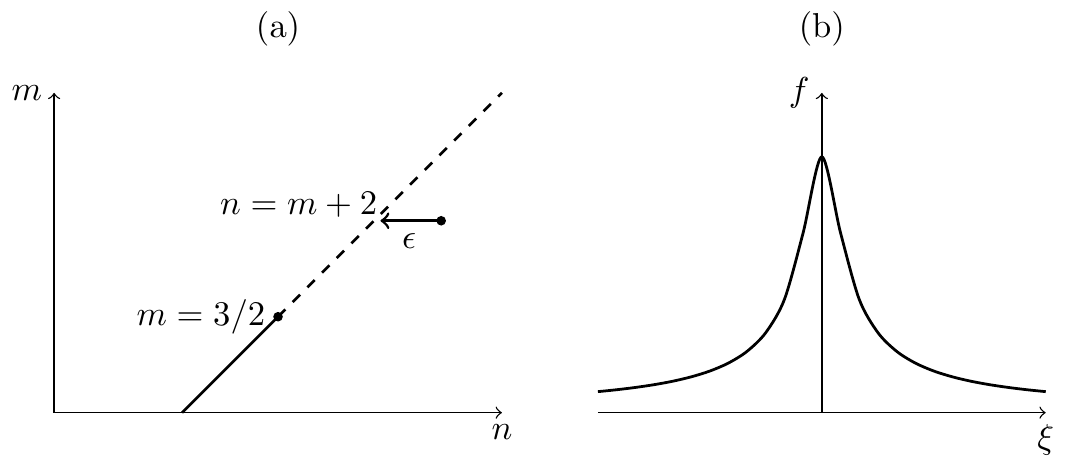}
\label{fig:schematic}
\caption{(a) The region of the $(m,n)$ plane under consideration: $n$ is greater than the critical line $n = m+2$ found by \cite{Bertozzi1998} by a small amount $\epsilon$, while $m$ is greater than the critical value $3/2$, above which self-similar solutions do not exist on the critical line. (b) A schematic single-peaked self-similar profile in the region under consideration, which represents a blow-up solution of (\ref{eq:pde}).}
\end{figure}

In this paper, however, we will focus on (\ref{eq:pde}) with sufficiently large values of $n$, for which the generic behaviour is finite-time blow up ($h(x_0,t) \to \infty$ as $t\to t_0^-$).
\citet{Bertozzi1998} showed that finite time blow up is impossible given periodic conditions when $n < m+2$, which defines a critical line in the $(m,n)$-plane (see figure \ref{fig:schematic}).  Although the proof by \cite{Bertozzi1998} is rigorous and makes no assumption that the blow up must be self similar, the value $n=m+2$ does have specific relevance for self-similar blow up; as the blow-up time is approached, the mass of the self-similar solution either grows without bound, is finite, or vanishes, for $n < m+2$, $n=m+2$, and $n>m+2$, respectively.  Thus, given that (\ref{eq:pde}) (with periodic or zero flux boundary conditions) conserves mass, self-similar blow up cannot occur when $n<m+2$.

On the critical line $n = m+2$ and when $m < 3/2$, there are a countable number of families of compactly supported self-similar blow-up solutions, where each solution is characterised by the region of compact support, and number of peaks~\cite{Slepcev2005,Witelski2004,Evans2007,Liu2017}.  These similarity solutions represent exact solutions of (\ref{eq:pde}) that draw all the mass into the singularity.  Blow up of the equation is only achieved if there is sufficient mass in the system; the critical mass goes to infinity as $m \rightarrow 3/2$ \cite{Slepcev2005}.

Less attention has been paid to supercritical values of $n$ ($n > m+2$).  \citet{Bertozzi2000} and \citet{Evans2006} extensively examine the cases $m=1$ and $m=0$, respectively.  However, our interest in this paper is values $m > 3/2$ (see figure \ref{fig:schematic}), as for these values (as we will see) there are self-similar solutions when $n>m+2$, although there are none where $n=m+2$.  In fact, the author's original motivation of studying this regime was the Benney equation (a first order thin film approximation formulated to model the effects of inertia on thin film stability), which is known to exhibit finite-time blow up and is approximated by (\ref{eq:pde}) with $m=3$ and $n=6$, sufficiently close to blow up \cite{Pumir1983}.

The paper is as follows.  In section \ref{sec:computation} we describe the formulation of the equation for self-similar blow-up profiles of (\ref{eq:pde}), as well as a useful rescaling and numerical computation.  The main result of the paper is in section \ref{sec:asymptotics} where we derive the matched asymptotic solution to the problem, finding the primary branch of blow-up profiles as $n \to m+2^+$.  This requires obtaining information about the first four terms in the asymptotic expansion in an outer region, as well as the leading order problem in an inner region (which usually must be determined numerically).  The asymptotic predictions are shown to be in agreement with the numerical results.  Finally we summarise in section \ref{sec:discussion}, and include further information in appendices on  the first correction to the outer problem, numerical computation of the inner problem, and a special case of an exact solution of the inner problem.

\section{Similarity solutions}
\label{sec:computation}

\subsection{Similarity equation}
Assuming a finite-time blow up occurs at $x=x_0$, $t=t_0$, we make the standard similarity ansatz:
\begin{equation}
\label{eq:ssansatz}
h(x,t) = (t_0-t)^{-\alpha} f(\xi), \qquad \xi = \frac{x-x_0}{(t_0-t)^\beta}.
\end{equation}
On substitution into the thin film equation (\ref{eq:pde}), and matching the powers of $(t_0-t)$ on each term, we find the similarity equation
\begin{equation}
\label{eq:ss}
\alpha f + \beta \xi \pfrac{f}{\xi} + \pfrac{}{\xi}\left[ f^m \pfrac{^3f}{\xi^3} + f^n \pfrac{f}{\xi} \right] = 0,
\end{equation}
along with the values for the similarity exponents $\alpha$ and $\beta$:
\begin{equation}
\label{eq:alphabeta}
\alpha = \frac{1}{2n-m}, \qquad \beta = \frac{n-m}{2(2n-m)}.
\end{equation}
The boundary conditions required for (\ref{eq:ss}) are that the first two terms in (\ref{eq:ss}) balance in the far field, thus
\begin{equation}
\label{eq:ff}
f \sim c_\pm |\xi|^{-\alpha/\beta}, \qquad \xi\rightarrow \pm\infty,
\end{equation}
where $\gamma_\pm$ are a priori unknown coefficients.  These are the so-called `quasi-stationary' conditions that specify that $h_t$ remain bounded away from the singularity at $x_0$ as $t\to t_0^-$ ~\cite{Zhang1999,Witelski1999}.  Similar to the rupture case, it can be shown that each far-field solution (\ref{eq:ff}) has a two-dimensional stable manifold as $\xi\to \pm\infty$ respectively, so the correct number of boundary conditions is attained.
We will assume the similarity profile is symmetric in $\xi$, so that $c_+ = c_-$ and the conditions at $\pm\infty$ can be replaced by
\begin{equation}
\label{eq:bc0}
f'(0) = f'''(0) = 0.
\end{equation}

Based on previous results for self-similar profiles on the critical line \cite{Evans2007} and for $m=0$ \cite{Evans2006} and $m=1$ \cite{Slepcev2005} it is almost certain that there will be a countably infinite number of self-similar branches, characterised by the number of peaks.  In this paper we will examine only the primary branch of single-peaked profiles; see the discussion in section \ref{sec:discussion}.

\subsection{Rescaling}

Numerical computation of (\ref{eq:ss}) with boundary conditions (\ref{eq:ff}), (\ref{eq:bc0}) reveals that the self-similar profiles become singular with the maximum $f_0 = f(0)$ tending to infinity as $n \to m+2^+$.  To clarify the asymptotic structure it is therefore advantageous to perform the following rescaling.  Let $f(\xi) = f_0\phi(z)$, $\xi = zf_0^{1+\epsilon/2}$, $n = m+2+\epsilon$, and define a small parameter $\delta(\epsilon) = f_0^{-(m+4+2\epsilon)/(2m-3)}$ (recall that we assume $m>3/2$).  This definition is chosen so that $\delta$ is the order of the inner region, as will be apparent later.  The similarity equation (\ref{eq:ss}) becomes
\begin{equation}
\label{eq:rescaledode}
\delta^{2m-3} (\alpha \phi + \beta z\phi') + (\phi^{m+2+\epsilon}\phi')' + (\phi^m\phi''')' = 0,
\end{equation}
where
\begin{equation}
\label{eq:alphabeta2}
\alpha = \frac{1}{m+4+2\epsilon}, \qquad \beta = \frac{2+\epsilon}{2(m+4+2\epsilon)}
\end{equation}
with boundary conditions
\begin{equation}
\label{eq:rescaledbcs}
\phi(0) = 1, \quad \phi'(0) = \phi'''(0) = 0, \quad \phi \sim \gamma z^{-\alpha/\beta}, \ z \to \infty,
\end{equation}
where $\gamma = \gamma(\epsilon)$.  Our ultimate aim is to find $\delta$ (asymptotically) as a function of $\epsilon$, and thus find the (primary) solution branch in the limit $\epsilon\to 0^+$, that is, $n\to m+2^+$.  We will refer to this as the \emph{selection problem}.

\subsection{Numerical computation}
To compute solutions to (\ref{eq:rescaledode}), (\ref{eq:rescaledbcs}) numerically, we use a similar method to that which was used in \cite{Dallaston2017,Dallaston2018} for self-similar rupture profiles.  A stable starting point is computed for $m=3,n=8$ using the time-dependent method described in the supplementary information in \cite{Dallaston2018}, and then rescaled to represent a solution to (\ref{eq:rescaledode}).   Solutions for other $(m,n)$ values are then found via numerical continuation using the software AUTO-07p \cite{Doedel2007}.  

\begin{figure}
\centering
\includegraphics{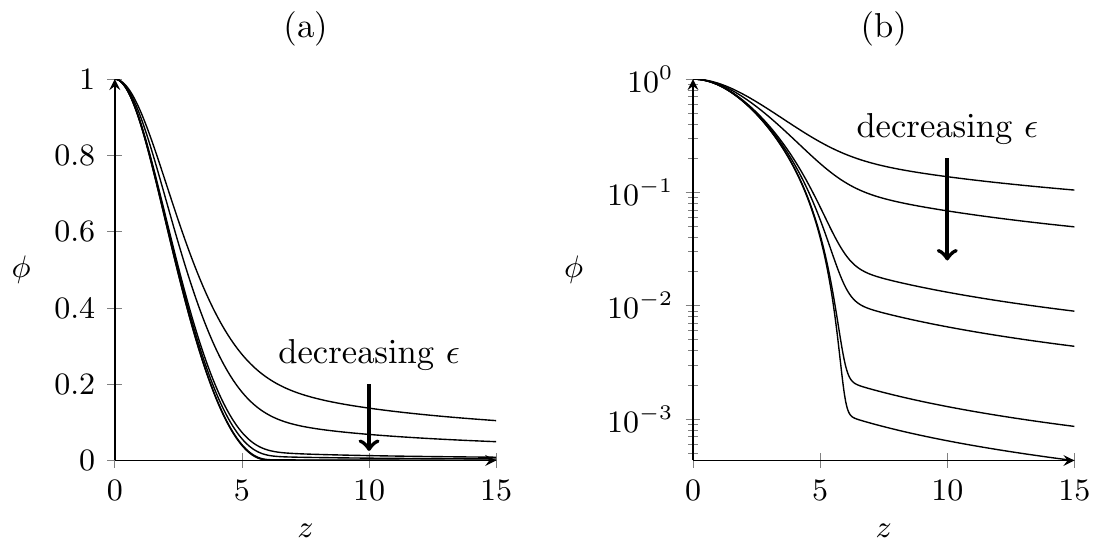}
\caption{(a) Numerically computed self-similar profiles (\ref{eq:rescaledode}) for $m=3$ and $\epsilon \in \{1,0.5,0.1,0.05,0.01,0.005\}$, showing the development of the asymptotic structure.  (b) The same solutions with $\phi$ on a logarithmic scale, revealing the scaling of the the inner and far field regions.}
\label{fig:sols}
\end{figure}

In figure \ref{fig:sols} we show solutions to (\ref{eq:rescaledode}), (\ref{eq:rescaledbcs}) for $m=3$, and for $\epsilon \in \{1,0.5,0.1,0.05,0.01,0.005\}$ (the picture is qualitatively similar for other values of $m$).  In this figure the asymptotic structure is apparent, with a single peak connected to a progressively thin far field region via an apparent `contact line', where the solution changes rapidly.  When we perform the matched asymptotic analysis, we will refer to these regions as the outer region, the far field region, and the inner region, respectively.

\begin{figure}
\centering
\includegraphics{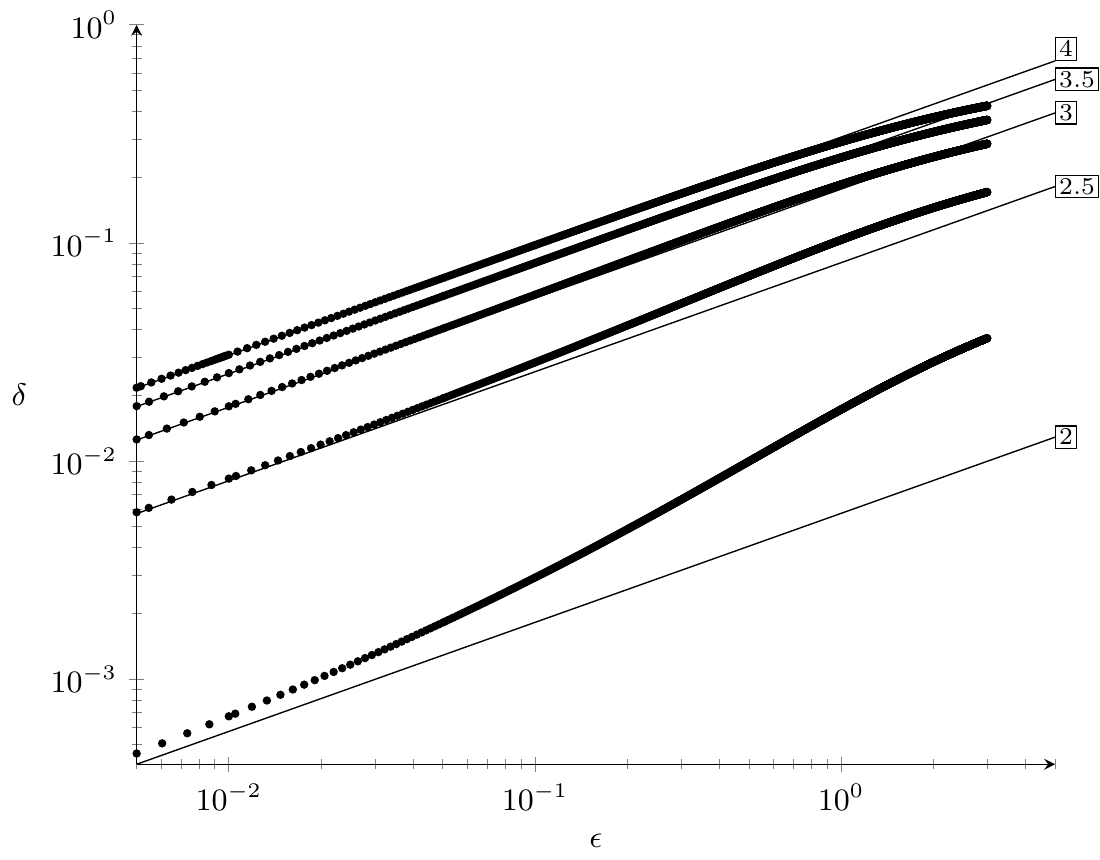}
\caption{The solution branches $\delta$ as a function of $\epsilon$, for $m \in \{2,5/2,3,7/2,4\}$.  The points represent numerical solutions using the method described in section \ref{sec:computation}, while the solid lines represent the predicted relationship from the matched asymptotic analysis (\ref{eq:k}). }
\label{fig:branches}
\end{figure}

As well as the solution profiles is is valuable to show the relationship between $\delta$ and $\epsilon$ as $\epsilon \to 0$.  In figure \ref{fig:branches} we plot the numerically evaluated points for $m \in \{2,5/2,3,7/2,4\}$ (along with the asymptotic predictions we will find in the next section).

\section{Matched asymptotics}
\label{sec:asymptotics}

\subsection{The outer region}
We now proceed to the matched asymptotics.  As suggested by the numerical results (figure \ref{fig:sols}), in the limit $\epsilon \to 0^+$, the profiles $\phi$ will have three distinct regimes:
\begin{enumerate}
\item[(a)] An outer region $0 \leq z < z^*$, near the peak of the profile, where $\phi = O(1)$ and $z = O(1)$;
\item[(b)] An inner region, where $z = z^* + O(\delta)$, $\phi=O(\delta^2)$, 
\item[(c)] A far field region $z > z^*$, where $z = O(1)$ and $\phi = O(\delta^2)$.
\end{enumerate}
To fully determine the first term in the $\delta$-$\epsilon$ relationship, it turns out that we need information about the corrections to the outer problem up to the $O(\delta^{2m-3} \epsilon)$ term.  Write  
\begin{equation}
\label{eq:phiexp}
\phi \sim \phi_0 + \epsilon \phi_1 + \ldots + \delta^{2m-3}\phi_2 + \ldots \epsilon\delta^{2m-3}\phi_3 + \ldots.
\end{equation}
Depending on the relationship between $\epsilon$ and $\delta$, as well as $m$, the ordering of the terms in (\ref{eq:phiexp}) may change, but the terms above ($\phi_j, j = 0,\ldots,3$),  are the ones that are important in the selection problem.

The leading order problem is
\[
(\phi_0^{m+2}\phi_0' + \phi_0^{m}\phi_0''')' = 0, \qquad \phi_0(0) = 1, \phi_0'(0) = \phi_0'''(0) = 0.
\]
This may be integrated three times to obtain
\begin{equation}
\label{eq:phi0sol0}
(\phi_0')^2 = \frac{1}{6}(\phi_0 - \phi_0^4),
\end{equation}
where a constant of integration has been determined by the left-hand boundary conditions and the requirement that $\phi = \phi' = 0$ at a point that will end up being the inner limit $z^*$.  The solution may thus be written implicitly as
\begin{equation}
\label{eq:phi0sol}
z = \int_{\phi_0}^1 \frac{\sqrt{6} \, \mathrm dp}{\sqrt{p-p^4}},
\end{equation}
and the value $z^*$ of $z$ at which $\phi_0 = \phi_0' = 0$ may be found by integration:
\begin{equation}
\label{eq:zs}
z^* = \int_0^1 \frac{\sqrt{6} \, \mathrm dp}{\sqrt{p-p^4}} \approx 5.94895.
\end{equation}
This leading order problem is the same as that found by \cite{Evans2007} for large $f_0$ in the critical case ($\epsilon = 0, m < 3/2$).  Unlike the present problem, in their case $\delta$ was free, characterising a one-parameter family of solutions.

For the matching, we require the behaviour of $\phi_0$ as the inner region is approached.  This is obtained from (\ref{eq:phi0sol0}) to be
\begin{equation}
\label{eq:phi0ff}
\phi_0 \sim \frac{1}{24}(z^*-z)^2+ O((z^*-z)^8), \qquad z\to {z^*}^-. 
\end{equation}

\subsection{Corrections to the outer problem}

We now examine the corrections in (\ref{eq:phiexp}).  While the solution to the $O(\epsilon)$ correction can be found explicitly (see appendix \ref{app:outercorrection}), there is little hope of obtaining explicit solutions for $\phi_2$ and $\phi_3$.  Fortunately, only the asymptotic behaviours of $\phi_1$, $\phi_2$, and $\phi_3$ as $z \to z^*$ are required to perform the matching, and each of these may be obtained analytically without resorting to the full solution.

The problem for $\phi_1$ may be integrated to give the equation
\begin{equation}
\label{phi1}
(\phi_1'' + \phi_0^2\phi_1)' = -\phi_0^2\log\phi_0 \phi_0', \quad \phi_1(0) = \phi_1'(0) = 0.
\end{equation}
(the condition $\phi_1'''(0)=0$ has been used to eliminate an integration constant).  There is one degree of freedom left in the outer problem, corresponding to (say) the right hand value:
\begin{equation}
\label{eq:phi1ff}
\phi_1 \sim \phi_1(z^*), \qquad z \to z^*.
\end{equation}  
In fact, the equation for $\phi_1$ can be integrated again to form a first order linear equation, which can then be solved in terms of an integrating factor, giving an explicit formula for $\phi_1$.  The details are in appendix \ref{app:outercorrection}, but are not needed to determine $\delta(\epsilon)$.

The equation for $\phi_2$ may also be integrated to give
\begin{equation}
\label{eq:phi2}
(\phi_2'' + \phi_0^2\phi_2)' = -\beta_0 z \phi_0^{1-m}, \qquad \phi_2(0) = \phi_2'(0) = 0,
\end{equation}
where $\beta_0 = \alpha_0 = 1/(m+4)$ is the leading order term in the $\epsilon$-expansion of the exponents $\beta$ and $\alpha$ from (\ref{eq:alphabeta2}). We now have little chance to produce an explicit solution, so instead purely focus on the behaviour near $z^*$.  In this limit, the right hand side of (\ref{eq:phi2}) induces the term
\begin{equation}
\label{eq:phi2ff}
\phi_2 \sim -\frac{\beta_0 z^*}{24^{1-m}(2m-5)(2m-4)(2m-3)}(z^*-z)^{5-2m}, \qquad z \to z^{*+}.
\end{equation}
(with logarithmic factors if $m=5/2$ or $m=2$, but this lines up with what occurs in the inner problem and does not change the results derived in this paper).

The equation for $\phi_3$ is more involved as it involves cross-terms of the first two corrections $\phi_1$ and $\phi_2$.  Here we only include the terms that are important for the matching.  Ultimately, the equation takes the form
\begin{equation}
\label{eq:phi3}
(\phi_0^m\phi_3''' + m\phi_0^{m-1}\phi_1\phi_2''' + \ldots)' = -\beta_0(z\phi_1)' - \alpha_1\phi_0 - \beta_1 z\phi_0', \quad \phi_3(0) = \phi_3'(0) = \phi_3'''(0) = 0,
\end{equation}
where the suppressed terms are equal to zero at both $z=0$ and at $z\to z^*$, so that they do not infuence the first term in the expansion near $z^*$.  In (\ref{eq:phi3}), $\alpha_1$ and $\beta_1$ are the $O(\epsilon)$ terms in the expansions of $\alpha$ and $\beta$ in (\ref{eq:alphabeta2}).  Integrating, and using the boundary conditions at $z=0$, we have
\begin{equation}
\label{eq:phi32}
\phi_0^m\phi_3''' + m\phi_0^{m-1}\phi_1\phi_2''' + \ldots = -\beta_0 z\phi_1 - \beta_1 z\phi_0 + (\beta_1-\alpha_1)\int_0^z \phi_0 \, \mathrm dz.
\end{equation}
From (\ref{eq:alphabeta2}), $\beta_1-\alpha_1 = 1/[2(m+4)] = \beta_0/2$, while $\phi_0^{m-1}\phi_2''' \sim -\beta_0z^*$, and
\[
\int_0^{z^*} \phi_0 \, \mathrm dz = \int_0^1 \frac{\sqrt 6 p \,\mathrm dp }{\sqrt{p-p^4}} = \sqrt{\frac{2}{3}}\pi.
\]
Putting this together with the farfield behaviours of $\phi_0$ (\ref{eq:phi0ff}) and $\phi_2$ (\ref{eq:phi2ff}) we find that the asymptotic behaviour of $\phi_3$ is
\begin{equation}
\label{eq:phi3ff}
\phi_3 \sim \frac{(m-1)\beta_0 z^*\phi_1(z^*) + \pi\beta_0/\sqrt 6}{24^{-m}(2m-3)(2m-2)(2m-1)}(z^*-z)^{3-2m}, \qquad z \to z^*.
\end{equation}

\subsection{The inner region}
The inner problem occurs near $z=z^*$, where $\phi$ becomes small and the dominant balance in (\ref{eq:rescaledode}) switches to the second and fourth terms.  Let $z = z^* + \delta t$ and $\phi(z) = \delta^2 w(t)$.  Then the equation is
\begin{equation}
\label{eq:w}
\alpha\delta w + \beta( z^* + \delta t)w' + \delta^{6+2\epsilon} (w^{m+2+\epsilon} w')' + (w^m w''')' = 0.
\end{equation}
In this region all we need is the leading order problem $w \sim w_0 + O(\delta)$, where $w_0$ satisfies \begin{equation}
\label{eq:w0}
w_0''' = -\frac{\beta_0 z^*(w_0 - w^*)}{w_0^m}, \qquad w_0 \sim a(-t)^2, \ t\to -\infty, \quad w_0 \to w^*, \ t\to\infty.
\end{equation}
The first condition ensures we can match to the outer solution (\ref{eq:phi0ff}) and the second condition ensures the solution will match to an appropriate far field condition.  Here $w^*$ is a constant of integration that must be determined by the matching.  Equation (\ref{eq:w0}) is a generalisation of the classic Landau--Levich `drag-out' problem \cite{Landau1942}.  In general, the solution (including the parameter $a$ above) must be determined by numerical computation, which will be described in section \ref{sec:compw}.  (\ref{eq:w0}) does have an exact solution in the special case $m=7/2$, however, which we describe further in section \ref{sec:exactSol} and appendix \ref{app:exactSol}.

\subsection{Far field region}
\label{sec:farfield}
The far field problem does not play a role in the selection of solutions (beyond requiring the leading order inner problem (\ref{eq:w0}) to tend to a constant as $t\to\infty$), but we describe it here for completeness.  In the far field, $w$ scales as in the inner problem, but the independent variable is $z$.  This is needed so that the far field problem can match with the inner problem at a finite $z = z^*$ and finite $w = w^*$:
\[
\alpha w + \beta z w' + \delta^{5+2\epsilon}(w^{m+2+\epsilon}w')' + \delta^3(w^mw''')' = 0.
\]
To leading order ($w = w_F + O(\epsilon)$) we obtain the simple equation
\[
\beta_0(w + zw') = 0, \qquad w(z^*) = w^*.
\]
Thus in the profile in the far field is 
\begin{equation}
w_F = \frac{z^*w^*}{z}.
\label{eq:wF}
\end{equation}
This also gives us the asymptotic formula for the far field coefficient $\gamma \sim \delta^{2}z^*w^*$, if desired.

\subsection{Matching}
We now perform the matching.  The relevant terms in the far-field ($t \to -\infty$) expansion of $w_0$ are
\begin{equation}
\label{eq:w0ff}
w_0 \sim a(-t)^2 + b(-t) + c + \tilde a(-t)^{5-2m} + \tilde b(-t)^{4-2m} + \tilde c(-t)^{3-2m}
\end{equation}
(as for the terms in the outer expansion, the exact ordering of these terms depends on $m$, but is not important for the selection problem).  Expressing (\ref{eq:w0ff}) in outer coordinates:
\begin{align}
\phi \sim a(z^*-z)^2 & + \delta b(z^*-z) + \delta^2c + \delta^{2m-3}\tilde a(z^*-z)^{5-2m} \notag \\
&  + \delta^{2m-2}\tilde b(z^*-z)^{4-2m} + \delta^{2m-1}\tilde c(z^*-z)^{3-2m}, \qquad z \to z^*.
\label{eq:innerouter}
\end{align}
For the gauge functions in (\ref{eq:innerouter}) to match each of (\ref{eq:phi0ff}), (\ref{eq:phi1ff}), (\ref{eq:phi2ff}), and (\ref{eq:phi3ff}), we require $\epsilon \sim k\delta^2$ for some coefficient $k$.  Matching at each order then gives
\begin{subequations}
\begin{align}
a &= \frac{1}{24}, \label{eq:matcha} \\
b &= 0, \label{eq:matchb} \\
c & = k\phi_1(z^*), \label{eq:matchc} \\
\tilde a &= -\frac{\beta_0 z^*}{24^{1-m}(2m-5)(2m-4)(2m-3)}, \label{eq:matchta} \\
\tilde b &= 0, \label{eq:matchtb} \\
\tilde c &= k\left[\frac{(m-1)\beta_0 z^*\phi_1(z^*) + \pi\beta_0/\sqrt 6}{24^{-m}(2m-3)(2m-2)(2m-1)}\right]. \label{eq:matchtc}
\end{align}
\end{subequations}
Given $b = \tilde b = 0$, direct expansion of $w_0$ in (\ref{eq:w0}) results in the following for  $\tilde a$ and $\tilde c$:
\begin{equation}
\label{eq:tildea}
\tilde a = -\frac{\beta_0z^*}{a^{m-1}(2m-5)(2m-4)(2m-3)}
\end{equation}
\begin{equation}
\label{eq:tildec}
\tilde c = \frac{\beta_0z^*[(m-1)c + w^*]}{a^m(2m-3)(2m-2)(2m-1)},
\end{equation}
while $a$ and $c$ (given $b=0$) must be found by computation of the generalised Landau--Levich problem (\ref{eq:w0}).

Now the matching (\ref{eq:matcha}) proceeds as follows: the leading order matching allows us to determine $w^*$ in the inner problem, the $O(\delta)$ matching (\ref{eq:matchb}) removes the degree of freedom in the inner problem to give $c$, the $O(\delta^{2m-3})$ and $O(\delta^{2m-2})$ matching (\ref{eq:matchta}, \ref{eq:matchtb}) are then identically satisfied, and the combined $O(\delta^2)$ and $O(\delta^{2m-1})$ matching (\ref{eq:matchc}), (\ref{eq:matchtc}) will give us both $\phi_1(z^*)$ and $k$.  In fact, fortuitous cancellation between (\ref{eq:tildec}) and (\ref{eq:matchtc}) mean that we find $k$ without having to compute $\phi_1(z^*)$ (and thus $c$); substituting (\ref{eq:matchc}) into (\ref{eq:matchtc}) and equating to (\ref{eq:tildec}), and rearranging for $k$, we find the remarkably simple formula
\begin{equation}
\label{eq:k}
\epsilon \sim k\delta^2, \qquad k = \frac{\sqrt 6 z^* w^*}{\pi}, \qquad \epsilon \to 0.
\end{equation}
In figure \ref{fig:branches} we show that this relation closely matches the numerical results in the small $\epsilon$ limit (once $w^*$ has been computed as below).  Of course, one may invert this relationship to find $\delta$, and thus $f_0$, as a function of $\epsilon$:
\[
f_0 \sim k^{(2m-3)/2(m+4)} \epsilon^{-(2m-3)/2(m+4)}, \qquad \epsilon \to 0.
\]
The only remaining issue in the selection problem is to find $w^*$, which requires numerical computation of the inner problem.  We describe this in the next section.

\subsection{Computation of $w^*$ and $c$}
\label{sec:compw}

In order to find $w^*$ for a given $m$, we rescale (\ref{eq:w0}) according to $W = w^* W$, and $t = (\beta z^*/w^{*m})^{-1/3}T$.  Then the inner problem is reduced to 
\begin{align}
\label{eq:W0}
W''' = -\frac{W - 1}{W^m}, \qquad W|_{T\to\infty} = 1, \notag \\
 W \sim A(-T)^2 + B(-T) + C, \qquad T\to-\infty,
\end{align}
where now $A$, $B$ and $C$ are functions of $m$ only, and
\begin{equation}
\label{eq:abcfromABC}
a = w^* \left(\frac{\beta_0 z^*}{w^{*m}}\right)^{2/3}A, \quad b = w^* \left(\frac{\beta_0 z^*}{w^{*m}}\right)^{1/3}B, \quad c = w^* C.
\end{equation}
Thus once we compute $A$, we immediately have $w^*$ from (\ref{eq:matcha}):
\begin{equation}
\label{eq:ws}
w^* = \left[\frac{24 A z^{*2/3}}{(m+4)^{2/3}}\right]^{3/(2m-3)}.
\end{equation}
The computation of $A$ is readily performed by extending on the method first used by \cite{Landau1942} who found $A$ for $m=3$ (see appendix \ref{app:computeABC}).  This is sufficient information to find the coefficient $k$ in (\ref{eq:k}), which we tabulate for some different values of $m$ in table \ref{tab:innerproblem}.

\begin{table}
\centering
\begin{tabular}{c|cccccc}
$m$ & $A$ & $D = B^2-4AC$ & $C (B=0)$ & $w^*$  & $c$ & $k$ \\ \hline
4& 0.2047 & -1.380 & 1.685 & 2.309  & 3.889 & 10.71 \\
7/2& 0.2500 & -2.000 & 2.000 & 3.414  & 6.827 & 15.83 \\
3& 0.3215 & -3.679 & 2.861 & 6.923  & 19.80 & 32.10 \\
5/2& 0.4526 & -2.307 & 1.274 & 32.75  & 41.74 & 151.9 \\
2& 0.7828 & 1.099 & -0.3510 & 6518.  & -2288. & 30228. \\ \hline
\end{tabular}
\caption{Numerically computed farfield coefficents in the generalised Landau--Levich problem (\ref{eq:W0}), and the resulting values of $w^*$ from (\ref{eq:ws}), $c$ from (\ref{eq:c}) and $k$ from (\ref{eq:k}).  The values of $A$ and $D$ for $m=7/2$ match those from the exact solution (\ref{eq:Wex}) to four significant figures.}
\label{tab:innerproblem}
\end{table}

Only $A$ is needed to determine the selection problem.  However, if we wanted to find the first correction term $\phi_1$ in (\ref{eq:phiexp}) we would need $\phi_1(z^*)$, and so would need to determine $c$, and thus, $C$.  Before matching, the solution to (\ref{eq:W0}) has one degree of freedom resulting from translational invariance; while $A$ is uniquely identified, the next terms $B$ and $C$ are related by the invariance of the discriminant-like quantity $D = B^2 - 4AC$.  The quantity $D$ can also be found numerically (see appendix \ref{app:computeABC}), and on setting $B=0$ from (\ref{eq:matchb}) this gives us $c$ as
\begin{equation}
\label{eq:c}
c = -\frac{w^* D}{4A}.
\end{equation}
These values are also given in table \ref{tab:innerproblem}.

\subsection{Exact solution to the inner problem for $m=7/2$}
\label{sec:exactSol}
It is interesting that although no closed form solution for the Landau--Levich problem is known for general $m$ (or for the hydrodynamically relevant value $m=3$), there is a closed form solution available for $m=7/2$ \cite{Polyanin2003}.  This solution is given implicitly by
\begin{equation}
\label{eq:Wex}
T = \log(\sqrt{W}+1) - \log(\sqrt{W}-1) - 2\sqrt W + K,
\end{equation}
where $K$ is arbitrary (see appendix \ref{app:exactSol} for details).  The $T\to-\infty$ expansion of $W$ here is
\[
W \sim \frac{1}{4}(-T)^2 - \frac{K}{2}(-T) + \frac{K^2+8}{4} 
\]
Thus $A = 1/4$ and $D = -2$.  This aligns with the values found numerically in table \ref{tab:innerproblem}.

\subsection{Comparison with numerical computation in each region}
\begin{figure}
\centering
\includegraphics{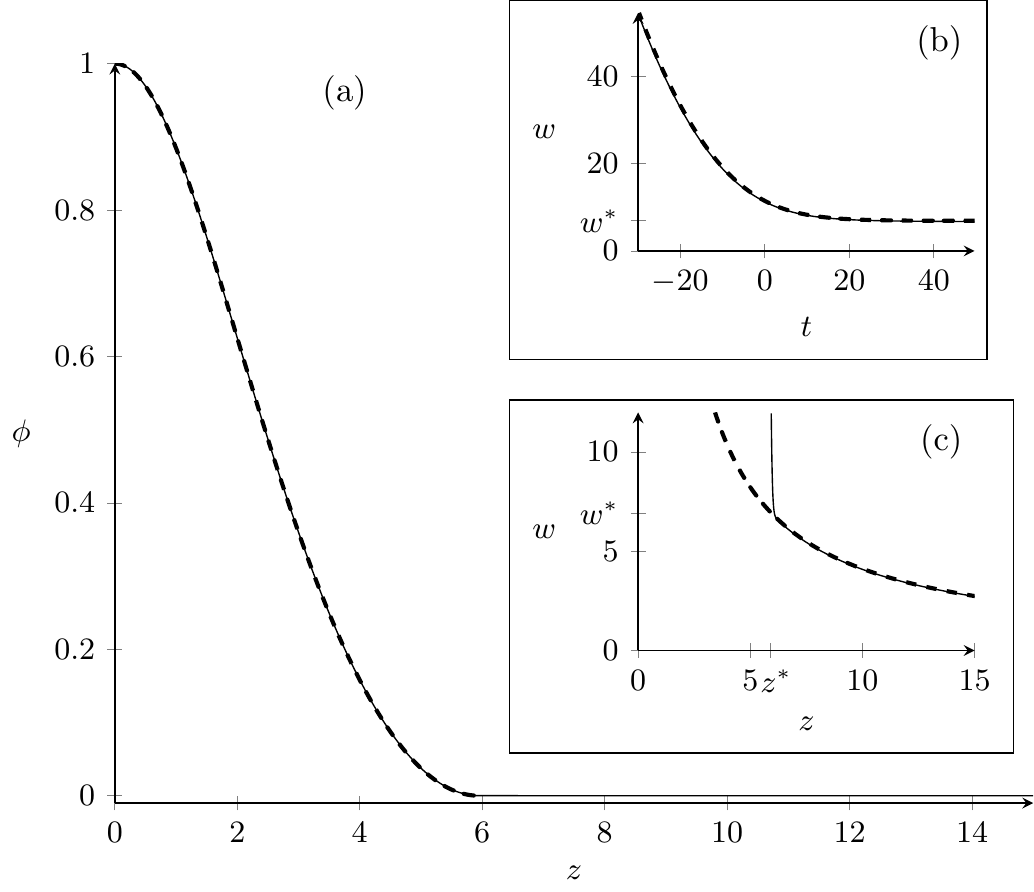}
\caption{Comparison between the leading order asymptotic results (dashed) and numerical results (solid) for each of the three asymptotic regions for $m=3$ and $\epsilon = 0.005$: (a) the outer region (\ref{eq:phi0sol}), (b) the inner region (\ref{eq:w0}), and (c) the far field region (\ref{eq:wF}).}
\label{fig:regions}
\end{figure}

In figure \ref{fig:regions} we plot the leading order solution in each region (\ref{eq:phi0sol}), (\ref{eq:w0}), (\ref{eq:wF}) against the numerical solution for $m=3$, $\epsilon = 0.005$.  This figure shows the validity of the solution in each region.
 
\section{Discussion}
\label{sec:discussion}

In this paper we have specifically examined the asymptotic structure of single-peaked self-similar blow-up profiles to the PDE (\ref{eq:pde}) in the region $m > 3/2$, $n = m+2 + \epsilon$, $\epsilon \to 0$.  Our result provides strong evidence that the critical line $n=m+2$ is sharp, and (\ref{eq:pde}) can indeed blow up for any value of $n > m+2$ where $m>3/2$.  

While we have computed the asymptotics of one branch of solutions to (\ref{eq:ss}), there are almost certainly an infinite number of such solutions, characterised by the number of peaks $J$, as found previously when  $n=m+2, m < 3/2$ \cite{Evans2007} (when there are an even number of peaks, $f_0$ has to be interpreted only as $\max(f)$, rather than $f(0)$).  The selection of such multi-peaked solutions is likely to be similar to that for the single peaked solution described in section \ref{sec:asymptotics}.  For $J$ peaks, the leading order outer solution $\phi_0$ consists of the even extension of (\ref{eq:phi0sol}), with period $2z^*$, and where the matching to the inner problem (\ref{eq:w0}) is applied at $z = Jz^*$.  Thus the integral on the left hand side of (\ref{eq:phi32}) will be multiplied by a factor of $J$, as will the factors of $z^*$ in (\ref{eq:phi32}) and (\ref{eq:tildec}), leading to the same formula for the correction problem (\ref{eq:k}).  Thus, we expect the multi-peaked branches of solutions to have asymptotically have the same height $f_0$ as the primary, single-peaked branch, for given $\epsilon$.  To confirm this numerically, a method of constructing the other branches would be required, for instance using a homotopy continuation method described in \cite{Tseluiko2013} for the rupture case.

The numerical quantities $w^*$, $c$ and $k$ of the inner problem blow up as $m \to 3/2^+$.  As this value is approached, the asymptotic approximation gets worse, as can be seen in figure \ref{fig:branches}.  If one takes $m < 3/2$ then the limit as $\epsilon \to 0$ is of a different nature, as then the solution branch will tend to one of the continuous family of solutions on the line $m=3/2$ without $f_0 \to \infty$ (see \cite{Bertozzi2000} for $m=1$, for instance).  This is a selection problem of a different sort, where small but finite $\epsilon$ picks one solution out of a continuum.

Finally, while the critical line $n=m+2$ below which blow up cannot occur has been known for some time (suggesting the asymptotic limit explored in this paper), the critical line above which finite time \emph{rupture} ($h\to 0$) cannot occur has not been determined; in that case the bifurcation structure of self-similar (and discretely self-similar) solutions is clearly more complicated \cite{Dallaston2018}, and it is not obvious whether a similar analysis could be done in that case.  Despite its ubiquity in thin film hydrodynamics, the equation (\ref{eq:pde}) clearly contains many critical values, and interesting asymptotic limits, in the $(m,n)$ plane that have yet to be fully explored.

\section*{Acknowledgment}

MCD acknowledges the generous support of the QJMAM Fund for Applied Mathematics for providing travel funding related to this research.


\appendix

\section{The correction to the outer problem}
\label{app:outercorrection}

The leading order solution to the outer problem is given implicitly by
\[
\phi_0' = -\frac{1}{\sqrt 6}\sqrt{\phi_0 - \phi_0^4}, \qquad \phi_0(0) = 1.
\]
The correction problem may be written
\[
\phi_1'' + \phi_0^2\phi_1 = \frac{\phi_0^3}{9} - \frac{\phi_0^3\log\phi}{3} + K_0
\]
where $K_0$ is an unknown constant.  Multiplying through by $\phi_0'$, using the leading order solution and integrating, we find
\[
\phi_0'\phi_1' - \phi_0''\phi_1 = \frac{7}{144}(\phi_0^4-\phi_0) - \frac{1}{12}\phi_0^4\log\phi_0 + K_1(\phi_0-1), \qquad K_1 = K_0 + \frac{7}{144}.
\]
The constant of integration has been determined from the left hand side vanishing at $z = 0$ ($\phi_0 = 1$).  Evaluating at $z = z^*$, we find $\phi_1(z^*) =12K_1$.

The above equation is first order and linear in $\phi_1$ with integrating factor $1/\phi_0'$.  Multiplying through by $1/\phi_0'^2$ and integrating again:
\begin{align*}
\phi_1 &= \phi_0'\int_a^z -\frac{7}{24} - \frac{1}{2}\frac{\phi_0^4\log\phi_0}{\phi_0 - \phi_0^4} - \frac{\phi_1(z^*)}{2} \frac{1-\phi_0}{\phi_0 - \phi_0^4} \, \mathrm dz\\
&= \sqrt{\phi_0 - \phi_0^4}\left[\frac{7}{24\sqrt{6}}z + \frac{1}{2}\int_{\phi_0}^1 \frac{p^4\log p}{(p-p^4)^{3/2}}\mathrm d p + \frac{\phi_1(z^*)}{2}\int_{\phi_0}^1 \frac{1-p}{(p-p^4)^{3/2}} \mathrm dp  \right]
\end{align*}
Here another arbitrary constant is fixed by requiring $\phi_1(0)=0$ (note each of the integrands are only weakly singular at $p=1$).  One parameter $\phi_1(z^*)$ remains in the solution, that has to be determined by the matching condition (\ref{eq:matchc}).

\section{Computation of coefficients of the inner problem}
\label{app:computeABC}

To compute $A$ and $D = B^2 - 4AC$ for general $m$ we extend on the method first used by \cite{Landau1942} who found $A$ for $m=3$.  Let $W$ be the independent variable and define $U = (W')^2$ to be the dependent variable.  Under these substitutions (\ref{eq:W0}) becomes
\begin{equation}
\label{eq:U0}
\tfrac{1}{2}\sqrt U U'' = \frac{1-W}{W^m}, \quad U|_{W\to 1} \to (W-1)^2, \ U'|_{W\to 1} \to 2(W-1).
\end{equation}
The nonlinear equation (\ref{eq:U0}) is readily solved numerically as an initial value problem from $W=1$.  

It is important to note the order of terms in the far field behaviour of $U$ changes as $m$ decreases.  When $m > 5/2$ the asymptotic behaviour of $U$ is 
\[
U = \hat A W + \hat D + \ldots
\]
and the coefficients may be easily extracted from the far field behaviour by taking $\hat A \approx U'(L)$ for large $L$, and $B \approx U(L) - \hat A L$.  These coefficients are then related to those in the far field expansion of $W(T)$ by
\[
A = \frac{\hat A}{4}, \qquad D = \hat D.
\]
When $2 < m \leq 5/2$, however, the term in the expansion induced by the $W^{1-m}$ term on the right hand side of (\ref{eq:W0}) is larger than the constant term and must be taken into account; for instance, for $m=5/2$ we have
\[
U = \hat A W + \frac{2}{\sqrt{\hat A}}\log W + \hat D + \ldots.
\]
In this case, it is most efficient to find $\hat A$ numerically as the appropriate root of $ U'(L) = \hat A + 2\hat A^{-1/2} W(L)^{-1}$ for $L\gg 1$, then subtract off the first two terms to find $\hat D$.

When $m\leq 2$ another term becomes larger than the constant term; for instance for $m=2$ we have
\[
U = \hat A W + \frac{8}{\sqrt{\hat A}} W^{1/2} - \frac{8}{\hat A^2}\log W + \hat D + \ldots
\]
Now we compute $\hat A$ as the appropriate root of $U'(L) = \hat A + 4\hat A^{-1/2}W^{-1/2} - 8\hat A^{-2} W(L)^{-1}$, and subtract off the first three terms to obtain $\hat D$.

\section{Exact solution for inner problem when $m=7/2$}
\label{app:exactSol}

The inner problem (\ref{eq:W0}) is
\[
W''' = \frac{1-W}{W^m}, \qquad m > 3/2.
\]
For $m=3$ this is the Landau--Levich equation that appears frequently as an inner problem in thin film flows connecting a thin liquid film to a thicker region.  No analytical solution is known in this case.

However, an exact solution can be constructed for the special value $m=7/2$ \cite{Polyanin2003}.  Perform the substitution:
\[
W = \frac{1}{\omega}, \qquad \frac{\mathrm d x}{\mathrm d\chi} = \frac{1}{\omega^{3/2}}.
\]
Then $W''' = -\omega^{5/2}\omega'''$ and we obtain
\[
\omega''' = \frac{1-\omega}{\omega^{7/2-m}}.
\]
When $m=7/2$ then, the equation is linear and easily solved:
\[
\omega = 1 + c_1\mathrm e^{-\chi} + c_2\mathrm e^{\chi/2}\sin\left(\frac{\sqrt 3}{2}\chi\right)
\]
(note the arbitrary shift in $\chi$ is not important).  In our case we seek the solution satisfying $\omega\to 1$ as $x\to\infty$ which corresponds to $c_2=0$, and we may set $c_1 = -1$ without loss of generality.  Then the solution is given parametrically by
\begin{align*}
W &= \frac{1}{1 - \mathrm e^{-\chi}}, \\
T &= \int\frac{\mathrm d\chi}{(1-\mathrm e^{-\chi})^{3/2}} = \log[\sqrt{1-\mathrm e^{-\chi}}+1] - \log[\sqrt{1-\mathrm e^{-\chi}}-1] - \frac{2}{\sqrt{1-\mathrm e^{-\chi}}} + K,
\end{align*}
with $\chi \in (0, \infty)$, where $K$ is a constant.  The solution may thus be written implicitly as:
\[
T = \log(\sqrt{W} + 1) - \log(\sqrt{W}-1) - 2\sqrt W + K, \qquad 1 < W < \infty.
\]
Note the above solution is equivalent to the solution of the second order problem (\ref{eq:U0}) where $m=7/2$:
\[
U = (W')^2 = \frac{(W-1)^2}{W}.
\]

\end{document}